\documentclass[aps,prb,reprint,showpacs,amsmath,amssymb,superscriptaddress,floatfix,longbibliography]{revtex4-1}
\usepackage{float}
\usepackage{amsfonts}
\usepackage{bm}
\usepackage{graphicx}

\begin{document}

\title{Edge states in non-Fermi liquids}

\author{I. V. Yurkevich}

\affiliation{Nonlinearity and Complexity Research Group, School of Engineering \& Applied Science, Aston University, Birmingham B4 7ET, United Kingdom}
%\affiliation{Institute of Fundamental and Frontier Sciences, University of Electronic Science and Technology of China, Chengdu 610054, People's Republic of China}

\begin{abstract}
We devise an approach to calculation of scaling dimensions of generic operators describing scattering within multi-channel Luttinger liquid. The local impurity scattering in arbitrary configuration of conducting and insulating channels is investigated and the problem is reduced to a single algebraic matrix equation. In particular, the solution to this equation is found for a finite array of chains described by Luttinger liquid models. It is found that for a weak inter-chain hybridisation and intra-channel electron-electron attraction the edge wires are robust against disorder whereas bulk wires, on contrary, become insulating. Thus, the edge state may exist in a finite sliding Luttinger liquid without time-reversal symmetry breaking (quantum Hall systems) or spin-orbit interaction (topological insulators).
\end{abstract}

\maketitle

\section{Introduction}
The recent advances in study of strongly correlated systems has led to wider search for exotic non-Fermi-liquid states in condensed matter systems. In particular, the quest for edge states protected against disorder has started. These states are protected by a symmetry that forbids perturbations potentially dangerous for the phase stability. The perturbations presenting a threat to stability may also be forbidden by a conservation law. Another option is the renormalisation of dangerous perturbations such that they become suppressed at low temperatures and vanish in zero-temperature regime leading to a 'protected' zero-temperature state. One of the promising models providing rich non-Fermi-liquid physics is an anisotropic system consisting of array of coupled one-dimensional wires. This model was used to for construction of integer \cite{Sondhi} and fractional quantum Hall states \cite{Kane2002}. Sliding phases in classical $XY$ models \cite{XY}, smectic metals \cite{smectic} and many other exotic states are all described by the sliding Luttinger liquid (sLL) model \cite{sLL}. Infinite arrays of wires and few-channel liquids have been investigated for decades. Their stability against disorder is another wide area of research. A single impurity embedded into a Luttinger liquid \cite{KF} and a continuous disorder in Luttinger liquid \cite{GS,BAA} are known to be essential for a single channel problem. The renormalisation group (RG) analysis shows that transport of repulsive electrons is completely blocked by a single impurity at zero temperature \cite{KF}.

The role of disorder (local impurity) embedded into a Luttinger liquid (LL) coupled to quasi-one-dimensional quadratic bath was explored in \cite{IVY}. Multi-channel strongly correlated systems with disorder have not been studied yet. The objective of this paper is to study local impurity coupled to a multi-channel Luttinger liquid. We will develop generic approach allowing treatment of various realisations of multi-channel Luttinger liquids where channels are either labeled by quantum numbers or spatially arranged to form an array of wires. In particular, we investigate effect of a single impurity onto array of weakly coupled Luttinger liquids (sliding Luttinger liquid). The results obtained also describe an array of wires with low density of multiple impurities at moderate temperatures when thermal length is smaller than the mean distance between impurities and, therefore, each impurity renormalises individually. We also assume that hybridisation, $t_{\perp}$, between channels/wires is small enough to ensure that the temperature, $T_{\perp}\sim D\,(t_{\perp}/D)^{\alpha}$ (where $D$ is the bandwidth and the exponent $\alpha=(2-\Delta_{\perp})^{-1}$ is defined through the scaling dimension, $\Delta_{\perp}$, of the most dangerous hybridization term) is low enough and cannot be reached. Under these assumptions we may safely use model of interacting but not hybridised wires (channels) with disorder modelled by individual impurities. And under these assumption we will demonstrate that in some region of parameters (intra- and inter-wire interaction strengths) we should observe edge states in 2D array of correlated wires without both magnetic field (preserved time-reversal symmetry) and spin-orbit scattering. The inter-wire interaction may enhance impurity scattering in the bulk wires (making them insulating) and suppress impurity scattering in the edge wires. This behaviour is similar to the one observed in quantum Hall systems or topological insulators. It is important to stress here, that the edge states we describe in this paper require neither time-reversal symmetry breaking nor spin-orbit interaction.

It is known that RG flow for a single channel problem depends on single parameter (so-called Luttinger parameter) K while excitations velocity does not play any role in the renormalisation \cite{KF}. Our general construction generalises this result: $N$-channel LL is described by $N$ velocities and a real symmetric $N\times N$ matrix which we call Luttinger ${\hat K}$-matrix responsible for impurity renormalisation. This matrix must be found from the algebraic matrix equation
\begin{equation}\label{K}
{\hat K}\,{\hat V}_+\,{\hat K}={\hat V}_-\,,
\end{equation}
where the matrices ${\hat V}_+$ and ${\hat V}_-$ describe density-density and current-current short-range (intra- and inter-channel) interactions between electrons. It is this 'Luttinger' ${\hat K}$-matrix that defines scaling dimensions of all possible scattering terms in all possible channel configurations (some are ideal, i.e conducting, and others insulating). Any process of $n_i$ backscattered (tunneled) particles in conducting (insulating) $i$-th channel is described by an operator with scaling dimension given by
\begin{equation}\label{Delta}
\Delta[{\bf n}]={\bf n}^{\rm T}\,{\hat\Delta}\,{\bf n}\,,\quad
{\hat\Delta}^{-1}={\hat{\cal P}}_{\rm c}\,{\hat K}\,{\hat{\cal P}}_{\rm c} + {\hat{\cal P}}_{\rm i}\,{\hat K}^{-1}\,{\hat{\cal P}}_{\rm i}\,,
\end{equation}
where ${\hat{\hat{\cal P}}}_{\rm c,i}$ are projectors onto conducting/insulating channels subspaces and ${\bf n}^{\rm T}=(n_1\,,...\,,n_N)$ is the integer-valued vector describing multiplicities of intra-channel scatterings. Just the fact that we may find all scaling dimensions from a single ${\hat K}$-matrix allows reconstruction of a generic phase diagram of a sLL. We will demonstrate how it works on two examples: mixture of two distinguishable $1D$ species and infinite array of wires (sLL). Our main result will be the application of this generic approach to a finite 2D array (strip) which will prove the existence the edge state robust against disorder (unlike bulk wires) for the weak intra-channel attraction between fermions.

We believe that our general approach, when applied to sLL, will enable reconstruction of generic phase diagram, provide simple scheme of calculating scaling dimensions of perturbations needed to analyse phase stability for the entire range of parameters, and allow identification of mixed states (described by RG unstable fixed points) corresponding to first-order phase transitions in such systems.

The paper is organized as follows. In Sec. II we recap multi-channel liquid description (the Green function and the Luttinger matrix are derived in the \ref{AsLL} and \ref{Vpm}), introduce description of arbitrary configuration consisting of a set of conducting and insulating channels, and construct Lagrangian capturing generic scattering by a local perturbation. Sec. III describes applications analysed earlier elsewhere simply to demonstrate how the devised approach works in the known situations (explicit calculations can be found the \ref{apps}). Sec. IV is devoted the problem of finite sLL in the nearest neighbours interaction approximation. And finally, in Sec. V we analyse the problem of the existence of the edge states that supports low-temperature transport when bulk is insulating. In \ref{SVP} and \ref{duality} one can find mapping of scaling dimension determination problem onto the Shortest Vector Problem (SVP) and the duality theorem which proves the absence of mixed 'inhomogeneous' phase with conducting edge mode (i.e. proving complete robustness of the phase with edge current).

\section{Multichannel Luttinger liquid Lagrangian}
Multichannel LL is a natural generalisation of one-dimensional LL. Different channels (further labeled by an integer $1\leq i\leq N$) might correspond to different quantum numbers (spin or quasi-spin nomenclature) or, if a spatial dimension added to the problem (geometry, i.e. 'distance' between channels), describe an anisotropic system composed of parallel chains with Luttinger liquids in each chain. All these problems allow universal description by assigning a density, $\rho_i$, and current, $j_i$, to the $i$-th channel (chain). Two vector fields
\begin{equation}
{\bm\theta}=(\theta_1\,,...\,,\theta_N)\,,\quad
{\bm\phi}=(\phi_1\,,...\,,\phi_N),
\end{equation}
are used to parameterise density and current in each channel:
\begin{equation}
\rho_i=\frac{1}{2\pi}\,\partial_x\theta_i\,, \quad
j_i=\frac{1}{2\pi}\,\partial_x\phi_i
\end{equation}
The propagation of these fields is governed by the Lagrangian
\begin{equation}\label{L0}
{\cal L}_0=\frac{1}{2}{\bm \Psi}^{\rm T}\,{\hat G}_0^{-1}\,{\bm\Psi}\,,\quad
{\hat G}_0^{-1}=\frac{1}{4\pi}\,\left[{\hat\tau}_1\,\partial_t+{\hat V}\,\partial_x\right]\,\partial_x\,,
\end{equation}
where matrix ${\hat V}$ is block-diagonal in density/current subspace and ${\hat\tau}_1$ is the Pauli matrix:
\begin{equation}%\label{G0}
{\bm \Psi}=\left(
              \begin{array}{c}
                {\bm\theta} \\
                {\bm\phi} \\
              \end{array}
            \right)\,,\quad
{\hat V}=\left(
              \begin{array}{cc}
                {\hat V}_+ & 0 \\
                0 & {\hat V}_- \\
              \end{array}
            \right)\,,\quad
{\hat\tau}_1=\left(
              \begin{array}{cc}
                0 & {\hat 1}\\
                {\hat 1} & 0 \\
              \end{array}
            \right)\,.
\end{equation}
The $N\times N$ matrix ${\hat V}_+$ (${\hat V}_-$) describes both intra- and inter-channel density-density (current-current) interactions. The ${\hat V}$-matrix is block-diagonal due to assumed inversion symmetry (${\bm \rho}\to{\bm \rho}$ and ${\bm j}\to -{\bm j}$).

The retarded Green function of translational invariant multi-channel LL (see \ref{AsLL})
\begin{equation}\label{GF0}
G_0^{\rm R}(x;\omega)=-\frac{2\pi i}{\omega}\left(
                                            \begin{array}{cc}
                                              {\hat K} & {\rm sgn}(x) \\
                                              {\rm sgn}(x) & {\hat K}^{-1} \\
                                            \end{array}
                                          \right)\,e^{i{\hat\omega}|x|}\,,
\end{equation}
where the Luttinger matrix ${\hat K}$ is found from the algebraic matrix Eq.~(\ref{K}).

It is instructive to notice here that the equal-coordinate (local) correlations are described by the effective (or 'local') Lagrangian, ${\cal L}_{\rm 0}^{\rm loc}$, similar to that used in \cite{KF} for a single channel problem. Zero-temperature description corresponds to the use of causal Green function obtained from the retarded component Eq.~(\ref{GF0}). The local Lagrangian acquires the form (see \ref{Vpm}):
\begin{eqnarray}
\hspace*{-30pt}
{\cal L}_{\rm 0}^{\rm loc}&=&i\int\frac{{\rm d}\omega}{2\pi}\,|\omega|\,{\bm \Psi}^{\dagger}_{\omega}\,
\left(
  \begin{array}{cc}
    {\hat K}^{-1} & 0 \\
    0 & {\hat K} \\
  \end{array}
\right){\bm \Psi}_{\omega}\,;\\
\quad {\bm \Psi}_{\omega}&=&\int{\rm d}t\,{\bm \Psi}(x=0,t)\,e^{i\omega t}.
\end{eqnarray}

\subsection{Local impurity}

Perturbing translation invariant system by a local impurity, one may find that the system is unstable in the sense that the scaling dimension of this perturbation is smaller than one (which is the physical dimension of a local term in $1+1$ dimensional system). This means that a perturbation theory would be divergent and allegedly translation invariant configuration is a bad initial approximation. In terms of transport, this fact suggests that some (all or few) of the channels (wires) become insulating and not ideally conducting as it is assumed in a translation invariant configuration. To test stability of such an inhomogeneous configuration where some channels are conducting while others are insulating we have to introduce two subspaces of conducting, ${\bf C}$, and insulating, ${\bf I}$, channels. Since we have no {\it \'{a} priori} knowledge which configuration will be stable (and, therefore, may be called a phase) against all meaningful perturbations, we must devise generic approach allowing treatment of all possible configurations.

The translation invariant channels, ${i\subset{\bf C}}$, are described by continuous $\theta_i$ - and $\phi_i$ -fields. The channels that, as an initial guess, are assumed to be insulating, ${i\subset{\bf I}}$, should be described by continuous $\theta_i$-fields which vanish at the position of interruption (at the origin $x=0$). The boundary conditions for insulating (disjoint at $x=0$) channel follow from the requirement that current (written as $j_i(x,t)=-\partial_t\theta_i(x,t)/2\pi$) at the tip of an insulating channel vanishes. Instead of using the boundary conditions described above, we will do a trick by adding to the translation invariant Lagrangian, ${\cal L}_0$, an auxiliary term
\begin{equation}
\hspace*{-20pt}
{\cal L}_{\xi}={\cal L}_0-\frac{1}{2}\,{\bm\theta}^{\rm T}\,{\hat\zeta}\,{\bm\theta}\,,\quad
{\hat\zeta}={\rm diag}(\zeta_1\,,...,\zeta_N)\,.
\end{equation}
The correlations found from the auxiliary Lagrangian ${\cal L}_{\xi}$ become true correlations of an inhomogeneous configuration after the limit
\begin{equation}\label{xi}
{\zeta}_i\,\to\,\left\{
\begin{array}{c}
0\,,\quad {i\subset{\bf C}}\,,\\
\infty\,,\quad {i\subset{\bf I}}\,.
\end{array}\right.
\end{equation}
The trick allows dealing with all configurations indifferently and the limit Eq.~(\ref{xi}), effectively suppressing currents in insulating channels (subset ${\bf I}$) and leaving unaffected conducting channels (subset ${\bf C}$), should be taken at the very end of calculations. In what follows, we will refer to this limiting procedure simply as $\xi$-limit (the name will become clear later when ${\zeta}$ will be renormalised and renamed to $\xi$).  The auxiliary Lagrangian involves the auxiliary Green function ${\hat G}_{\xi}$
\begin{equation}
{\cal L}_{\xi}=\frac{1}{2}{\bm \Psi}^{\rm T}\,{\hat G}_{\xi}^{-1}\,{\bm\Psi}\,,
\end{equation}
containing 'boundary term'
\begin{equation}
{\hat G}_{\xi}^{-1}={\hat G}_0^{-1}-{\hat\Sigma}\,,\quad {\hat\Sigma}(x,x')=\left(
                                                                              \begin{array}{cc}
                                                                                {\hat\zeta} & 0 \\
                                                                                0 & 0 \\
                                                                              \end{array}
                                                                            \right)\,\delta(x)\delta(x')\,.
\end{equation}
To calculate scaling dimensions of scattering operators one needs correlations
\begin{equation}\label{GF}
\langle{\bm \Psi}(x,t)\otimes{\bm \Psi}^{\rm T}(x',t')\rangle = iG_{\xi}(x,x';t,t')\,.
\end{equation}
The Green function is defined on real time axis for zero temperature or on the Keldysh contour for a finite temperature. The Green function can be found with the use of scattering ${\hat T}$-matrix. The equation for retarded Green function in ($\omega\,,x$)-representation is given by
\begin{equation}
{\hat G}_{\xi}^{\rm R}(x,x';\omega)={\hat G_{0}}^{\rm R}(x-x';\omega)+{\hat G_{0}}^{\rm R}(x;\omega)\,{\rm T}_{\xi}^{\rm R}(\omega)\,{\hat G_{0}}^{\rm R}(-x';\omega)\,,
\end{equation}
where ${\hat T}$-matrix can be expressed through the (local) self-energy ${\hat{\Sigma}}$, generated by the boundary term
\begin{equation}
{\hat T}_{\xi}^{\rm R}(\omega)={\hat \Sigma}^{\rm R}\,\left[1-{\hat G}_0^{\rm R}(0;\omega){\hat \Sigma}^{\rm R}\right]^{-1},\,\,\,
{\hat \Sigma}^{\rm R}=
\left(
\begin{array}{cc}
 {\hat\zeta} & 0 \\
  0 & 0 \\
  \end{array}
  \right)\,.
\end{equation}

Now we can calculate the ${\hat T}$-matrix
\begin{equation}
{\hat T}_{\xi}=\frac{\omega}{2\pi i}\left(
           \begin{array}{cc}
             {\hat R}_{\xi} & 0 \\
             0 & 0 \\
           \end{array}
         \right)\,,\,\,\,
{\hat R}_{\xi}=\frac{1}{{\hat K}+{\hat\xi}^{-1}}\,,\,\,\,
{\hat\xi}=\frac{2\pi i}{\omega}\,{\hat\zeta}\,,
\end{equation}
introducing dimensionless auxiliary diagonal matrix ${\hat\xi}$ which justifies previously used notation for $\xi$-limit.
The Green function (before the $\xi$-limit) can be written as
\begin{eqnarray}\label{G}
{\hat G}_{\xi}&=&-\frac{2\pi i}{\omega}\,
\left[\left(
  \begin{array}{cc}
    {\hat K} & s_{x-x'} \\
    {\rm s}_{x-x'} & {\hat K}^{-1} \\
  \end{array}
\right)e^{i{\hat\omega}|x-x'|}\right.\\\nonumber
&-&\left.\left(
  \begin{array}{cc}
    {\hat K}\,{\hat R}_{\xi}\,{\hat K} & {\hat K}\,{\hat R}_{\xi}\,s_{-x'} \\
    {\rm s}_{x}\,{\hat R}_{\xi}\,{\hat K} & {\rm s}_{x}\,{\hat R}_{\xi}\,s_{-x'} \\
  \end{array}
\right)e^{i{\hat\omega}(|x|+|x'|)}\right]\,,
\end{eqnarray}
where the notation ${\rm s}_{x}={\rm sgn}(x)$ has been used. Since calculation of scaling dimension of a local perturbation requires only asymptotes of the correlations between fields at the origin $(x=\pm 0, x'=\pm 0)$ we may use the 'local' limit of Eq.~(\ref{G}):
\begin{eqnarray}\label{Gloc}
{\hat G}_{\xi}(x,x';\omega)\underset{x,x'\rightarrow 0}{\rightarrow}-\frac{2\pi i}{\omega}\,\times\\\nonumber
\left(\begin{array}{cc}
   (1-{\hat K}{\hat R}_{\xi}){\hat K} & s_{x-x'}+{\hat K}\,{\hat R}_{\xi}\,s_{x'} \\
    {\rm s}_{x-x'}-{\rm s}_{x}\,{\hat R}_{\xi}\,{\hat K} & {\hat K}^{-1}+{\hat R}_{\xi}\,s_{xx'}\\
\end{array}\right)\,.
\end{eqnarray}

\subsection{Scaling dimension of the scattering term}
The most general term describing process of simultaneous backscattering of few particles in the conducting channels and accompanying it transfer (tunneling) of particles across the cut in the insulating channels is written as
\begin{equation}\label{pert}
{\mathcal L}_{\rm pert}=\sum_{\bf n}{\mathcal L}_{\bf n}\,,\quad {\mathcal L}_{\bf n}=v_{\bf n}\,e^{i{\bf n}^{\rm T}\,{\bm\Phi}}+\mathrm{c.c.}\,,
\end{equation}
where the fields are defined through their projections onto (as it is guessed initially) conducting and insulating channels subspaces
\begin{equation}
{\bm\theta}_{\rm {\rm c}}={\hat{\cal P}}_{\rm c}\,{\bm\theta}\,,\quad {\bm\varphi}_{\rm {\rm i}}={\hat{\cal P}}_{\rm i}\,{\bm\varphi}\,,\quad
{\bm\Phi}={\bm\theta}_{\rm c}+\,{\bm\varphi}_{\rm i}\,.
\end{equation}
The new field ${\bm\varphi}$ describes the old $\phi$-field discontinuity around the scatterer (placed at the origin):
\begin{equation}\label{disc}
{\bm\varphi}=\frac{1}{2}\left[{\bm \phi}(x=+0)-{\bm \phi}(x=-0)\right]\,.
\end{equation}
We have also defined integer-valued vector ${\bf n}^{\mathrm{T}}=(n_1, n_2,\,...\,,n_N)$ that counts multiplicity of particle scattered by impurity.
There are two different processes that appear in the scattering term Eq.~(\ref{pert}). First, it is backscattering in conducting channels where only ${\bm\theta}$-fields from conducting channels, ${\bm\theta}_{\rm c}$, define this process. Second, the tunneling in insulating channels occurs and it is governed by the fields discontinuity Eq.~(\ref{disc}).

The Eq.~(\ref{pert}) represents the most general perturbation applied to an arbitrary configuration if we neglect (as it was discussed in the introduction) scattering accompanied by a change of the channel index. This is justified when conservation laws prohibit change of the quantum number corresponding to the nomenclature of channels or when overlap between wavefunctions belonging to different wires (channels) is small. But we would like to stress that we restrict ourselves to a perturbation given above only to keep it simple. If one needs to account for a specific scattering process, the only thing which has to be done is relabeling of {\it chiral} channels to make sure that the important scattering event occurs inside the same non-chiral channel. Such a procedure will corrupt the matrices ${\hat V}_{\pm}$ but otherwise the whole approach will be intact.

Not all correlations present in Eq.~(\ref{Gloc}) are needed for calculation of scaling dimensions of the scattering terms.
All the necessary for this purpose correlations are contained in the reduced Green function:
\begin{equation}
i\,{\cal G}_{\xi}=\left(
\begin{array}{cc}
\langle{\bm\theta}_{\rm c}\otimes{\bm\theta}_{\rm c}^{\rm T}\rangle & \langle{\bm\theta}_{\rm c}\otimes{\bm\varphi}_{\rm i}^{\rm T}\rangle \\
\langle{\bm\varphi}_{\rm i}\otimes{\bm\theta}_{\rm c}^{\rm T}\rangle & \langle{\bm\varphi}_{\rm i}\otimes{\bm\varphi}_{\rm i}^{\rm T}\rangle \\
\end{array}\right)\,.
\end{equation}
The retarded reduced Green function is readily extracted from Eq.~(\ref{Gloc}) in the form ${\cal G}_{\xi}=-(2\pi\,i/\omega)\,{\hat Q}_{\xi}$ with
\begin{equation}
{\hat Q}_{\xi}=\left(
  \begin{array}{cc}
    {\hat{\cal P}}_{\rm c}\,\left[{\hat K}^{-1}+{\hat\xi}\right]^{-1}\,{\hat{\cal P}}_{\rm c} & {\hat q}_{\xi} \\
    -{\hat q}^{\rm T}_{\xi} & {\hat{\cal P}}_{\rm i}\,\left[{\hat K}+{\hat\xi}^{-1}\right]^{-1}\,{\hat{\cal P}}_{\rm i} \\
  \end{array}
\right)\,,
\end{equation}
where the off-diagonal elements are given by the following expression:
\begin{equation}
{\hat q}_{\xi}={\hat{\cal P}}_{\rm c}\,{\hat K}\,\left[{\hat K}+{\hat\xi}^{-1}\right]^{-1}\,{\hat{\cal P}}_{\rm i}\,.
\end{equation}
And finally, we have to perform the $\xi$-limit. It is convenient to do it after presenting ${\hat K}$-matrix in the block form distinguishing conducting and insulating subspaces of channels:
\begin{equation}\label{block}
{\hat K}=\left(
             \begin{array}{cc}
               {\hat K}_{\rm cc} & {\hat K}_{\rm ci} \\
               {\hat K}_{\rm ic} & {\hat K}_{\rm ii} \\
             \end{array}
           \right)\,,\quad
{\hat K}^{-1}=\left(
             \begin{array}{cc}
               {\hat{\bar K}}_{\rm cc} & {\hat{\bar K}}_{\rm ci} \\
               {\hat{\bar K}}_{\rm ic} & {\hat{\bar K}}_{\rm ii} \\
             \end{array}
           \right)\,,
\end{equation}
with blocks defined by the projectors
$$
{\hat K}_{\mu\nu}= {\hat{\cal P}}_{\mu}\,{\hat K}\,{\hat{\cal P}}_{\nu}\,, \quad
{\hat {\bar K}}_{\mu\nu}= {\hat{\cal P}}_{\mu}\,{\hat K}^{-1}\,{\hat{\cal P}}_{\nu}\,,\quad \mu,\nu={\rm {\rm c}, {\rm i}}\,.
$$
The $\xi$-limit brings the final result for the correlations needed to find scaling dimensions of the perturbations Eq.~(\ref{pert}) into the form:
\begin{equation}
\lim_{\xi}\,{\cal G}_{\xi}\equiv {\cal G}=-\frac{2\pi\,i}{\omega}\,
\left(
  \begin{array}{cc}
    {\hat{\bar K}}_{\rm cc}^{-1} & {\hat q}\\
    -{\hat q}^{\rm T}\, & {\hat K}_{\rm ii}^{-1} \\
  \end{array}
\right)\,,\quad
{\hat q}={\hat K}_{\rm ci}\,{\hat K}_{\rm ii}^{-1}\,.
\end{equation}

Using results from the previous section, we can write correlations in the time domain (these are actually greater or lesser Green functions with the corresponding infinitesimal imaginary shift which is not written explicitly below):
\begin{eqnarray}\label{corr}
%\begin{array}{cc}
\langle{\bm\theta}_{\rm c}(t)\otimes{\bm\theta}_{\rm c}^{\mathrm{T}}(t')\rangle &=-2{\hat{\bar K}}_{\rm cc}^{-1}\,\ln(t-t')\,,\\\nonumber
\langle{\bm\varphi}_{\rm i}(t)\otimes{\bm\varphi}_{{\rm i}}^{\mathrm{T}}(t')\rangle &= -2{\hat K}_{\rm ii}^{-1}\,\ln(t-t')\,.
%\end{array}
\end{eqnarray}
There are also cross-correlations (unimportant for our purposes as it will be seen later):
\begin{eqnarray}\label{cross}
\langle{\bm\theta}_{\rm c}(t)\otimes{\bm\varphi}_{\rm i}^{\mathrm{T}}(t')\rangle=-2{\hat q}\,\ln(t-t')\,,\\\nonumber
\langle{\bm\varphi}_{\rm i}(t)\otimes{\bm\theta}_{{\rm c}}^{\mathrm{T}}(t')\rangle=2{\hat q}^{\mathrm{T}}\,\ln(t-t')\,.
\end{eqnarray}
The scaling dimensions, $\Delta[{\bf n}]$, of exponentials in Eq.~(\ref{pert}) are defined by the correlations:
\begin{eqnarray}\label{-T}
\langle e^{i{\bf n}^{\rm T}\,{\bm\Phi}(t)}\,e^{-i{\bf n}^{\rm T}\,{\bm\Phi}(t')}\rangle &\sim & \\\nonumber
\, \exp\left[{\bf n}^{\rm T}\,\langle{\bm\Phi}(t)\otimes{\bm\Phi}^{\mathrm{T}}(t')\rangle{\bf n}\right]
&=&(t-t')^{-2\Delta[{\bf n}]}\,.
\end{eqnarray}
Since cross-correlations in Eq.~(\ref{cross}) are transposed to each other with the minus sign, they do not contribute to the correlation function defining scaling dimension in Eq.~(\ref{-T}). Using Eqs.~(\ref{corr}), one can write down scaling dimensions in the following form:
\begin{equation}
\mathrm{dim}[\mathcal{L}_{\bf n}]=\Delta[\bf n]={\bf n}_{\rm c}^{\mathrm{T}}\,{\hat{\bar K}}_{\rm cc}^{-1}\,{\bf n}_{\rm c}
+{\bf n}_{\rm i}^{\mathrm{T}}{\hat K}_{\rm ii}^{-1}\,{\bf n}_{\rm i}\,,
\end{equation}
where ${\bf n}_{\rm c}={\mathcal{P}}_{\rm c}\,{\bf n}\,,\, {\bf n}_{\rm i}={\mathcal{P}}_{\rm i}\,{\bf n}$. Finally, using properties of the projectors the scaling dimension can finally be written in the form presented earlier in the Introduction, Eq.~(\ref{Delta}):
\begin{equation}
\Delta[{\bf n}]={\bf n}^{\rm T}\,{\hat\Delta}\,{\bf n}\,,\quad
{\hat\Delta}^{-1}=\mathcal{P}_{\rm c}\,{\hat K}^{-1}\,\mathcal{P}_{\rm c}
+{\mathcal{P}}_{\rm i}\,{\hat K}\,{\mathcal{P}}_{\rm i}\,.
\end{equation}

\section{Applications}
Now the problem of finding scaling dimensions of scattering operators is reduced to the algebraic matrix equation Eq.~(\ref{K}) giving the Luttinger ${\hat K}$-matrix which should then be plugged into Eq.~(\ref{Delta}). It can be easily done in two rather simple limits that have been treated earlier \cite{KLY,smectic} otherwise.

\subsection{Two-channel and sLL models}
In the case of two channels, the matrix Eq.(\ref{K}) can be solved exactly for arbitrary density-density and current-current intra- and inter-channel interactions. This model was fully analysed in \cite{KLY}. The \ref{apps} describes the method of solving this nonlinear matrix equation with the use of the Cayley-Hamilton theorem. Of course, spinful LL model \cite{KF} is a limiting case that corresponds to the identical channels (equal velocities and intra-channel interactions) and is easily recovered from the Eqs.~(\ref{K},\ref{Delta}).

Infinite system of wires arranged into ordered arrays (sliding Luttinger liquid) was analysed long ago \cite{sLL}. We reproduce their results in \ref{sLL} from the solution of our main equations Eqs.(\ref{K},\ref{Delta}) just to demonstrate how the developed approach works in this situation.\\

There are also situations when a ${\it finite}$ array of spatially ordered wires can be treated exactly. The importance of this situation lies in the fact that we may analyse boundary of sLL (surface or edge states depending of spatial arrangements). Below we will show that a solution of the Eq.~(\ref{K}) can be obtained for a finite strip in the nearest neighbours interaction case. This allows application of the scheme devised in this paper and, therefore, analysis of the edge wire stability against impurity scattering. As one will see later, under some conditions the edge wires are robust against impurity scattering whereas the bulk wires are blocked by impurities. This corresponds to situation typical for topological insulators where bulk is insulating and edge is conducting.

\section{Finite array}
The only difference between channels being quantum numbers and wires arranged into arrays is the spatial dimension which defines matrices ${\hat V}_{\pm}$. In the case of channels thought of as quantum numbers, these matrices are featureless. For wires arranged into arrays the interaction should decay with the distance and this imposes structure onto matrices ${\hat V}_{\pm}$ and, therefore, the Luttinger ${\hat K}$-matrix. The most dramatic situation is the nearest neighbours interaction which correspond to symmetric tridiagonal ${\hat V}_{\pm}$ matrices:
\begin{equation}
V_{\pm}^{ij}=v_{\pm}\,\delta_{ij}+v_{\pm}'\,\left[\delta_{i,j+1}+\delta_{i,j-1}\right]\,.
\end{equation}
Here $v_{\pm}'$ describe inter-channel density-density and current-current interactions while $v_{\pm}$ encode intra-channel properties. Without inter-channel interaction we would have two noninteracting channels (wires) described by velocity $v$ and Luttinger parameter $K$:
\begin{equation}
v=\sqrt{v_+\,v_-}\,, \quad K=\sqrt{\frac{v_-}{v_+}}\,.
\end{equation}
Under inter-channel interaction one has to calculate the Luttinger matrix ${\hat K}$ instead of a single Luttinger parameter $K$.
All symmetric tridiagonal matrices are diagonalised by the same orthogonal, ${\hat{\cal O}}\,{\hat{\cal O}}^{\rm T}=1$, similarity transformation
\begin{eqnarray}
{\hat V}_{\pm}={\hat{\cal O}}\,{\hat V}^{\rm d}_{\pm}\,{\hat{\cal O}}^{\rm T}\,,\quad
{\cal O}_{ij}=\sqrt{\frac{2}{N+1}}\,\sin \frac{\pi ij}{N+1}\,.
\end{eqnarray}
The eigenvalues of matrices ${\hat V}_{\pm}$ are given by:
\begin{equation}
{\hat V}^{\rm d}_{\pm}={\rm diag}\left(V_{\pm}^1\,,...\,,V_{\pm}^N\right)\,,\quad
V^k_{\pm}=v_{\pm}+2v_{\pm}'\,\cos \frac{\pi k}{N+1}\,.
\end{equation}
for $k=1\,,...\,,N$. Since both interaction matrices are diagonalised by the same orthogonal transformation, ${\hat K}={\hat{\cal O}}\,{\hat K}^{\rm d}\,{\hat{\cal O}}^{\rm T}$, the $K$-matrix becomes also diagonal after this transformation as follows from the Eq.~(\ref{K}):
\begin{equation}
{\hat K}^{\rm d}={\rm diag}\left[K_1\,,...\,K_N\right]\,,\quad K_k=\sqrt{\frac{V_{-}^k}{V_{+}^k}}\,.
\end{equation}
The matrix elements of the ${\hat K}$-matrix and its inverse are then given by
\begin{eqnarray}
\hspace*{-20pt}
K_{ij}&=&\frac{2}{N+1}\,\sum\limits_{k=1}^{N}\,K_k\,\sin \frac{\pi ki}{N+1}\,\sin \frac{\pi kj}{N+1}\,,\\
\left[K^{-1}\right]_{ij}&=&\frac{2}{N+1}\,\sum\limits_{k=1}^{N}\,K^{-1}_k\,\sin \frac{\pi ki}{N+1}\,\sin \frac{\pi kj}{N+1}\,.
\end{eqnarray}

The notion of the edge state is well defined in a wide array with $N\gg 1$. In this limit the matrix elements of the ${\hat K}$-matrix factorize:
\begin{equation}
{\hat K}=K\,{\hat \kappa}\,,\quad \kappa_{ij}=2\,\int\limits_{0}^{\pi}\frac{{\rm d}q}{\pi}\,\kappa_q\,\sin qi\,\sin qj\,,
\end{equation}
where $K$ is the intra-channel Luttinger parameter whereas matrix ${\hat\kappa}$ contains renormalisations due to inter-channel interactions
\begin{equation}
\kappa_q=\sqrt{\frac{1+\alpha_-\,\cos q}{1+\alpha_+\,\cos q}}\,,\quad
\alpha_{\pm}=\frac{2\,v_{\pm}'}{v_{\pm}}\,.
\end{equation}

\section{Edge state}
In the regime of weak inter-channel overlap we may neglect impurity scattering resulting in particle changing the channel. Complete analysis of phase diagram will be reported elsewhere. We are currently interested in the situation when bulk is insulating due to relevant intra-channel reflections by impurities. To test whether such a state is stable against perturbations (i.e. exist as a phase) we have to find scaling dimensions of intra-channel tunneling operators and ensure that all tunneling events are irrelevant.

\begin{figure}
\centering
%\begin{minipage}{0.4\textwidth}
%\centering
%\includegraphics[width=.6\linewidth]{K_ii.pdf}
%\captionof{figure}{The phase diagram.}
%\label{cross-section}
%\end{minipage}%
%\quad\begin{minipage}{0.4\textwidth}
%\centering
\includegraphics[width=\columnwidth]{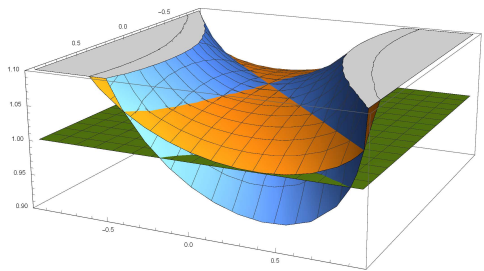}
\caption{The scaling dimensions (vertical axis) of bulk (blue surface) and edge (yellow surface) perturbations as functions of inter-wire interaction parameters $\alpha_+$ and $\alpha_-$ (horizontal axes). The region where the blue surface is under and the yellow surface is above the green plane falls into the region defined by Eq.~(\ref{ti}) and corresponds to the phase with insulating bulk and conducting edge state.}
\label{3D}
%\end{minipage}
\end{figure}

An impurity creating one-particle intra-channel backscattering in the $i$-th channel is described by the perturbation (see Eq.~\ref{pert}) with the scaling dimension $\Delta_i$ given by
\begin{equation}
\Delta_{ii}=\frac{2}{K}\,\int\limits_{0}^{\pi}\frac{{\rm d}q}{\pi}\,\kappa_q\,\sin^2 qi\,.
\end{equation}
In the bulk ($i\gg 1$) and at the edge ($i=1$) the backscattering terms have the following scaling dimensions:
\begin{eqnarray}
\Delta_{\rm bulk}&=&K\,\int\limits_{0}^{\pi}\frac{{\rm d}q}{\pi}\,\kappa_q\,,\\
\Delta_{\rm edge}&=&2K\,\int\limits_{0}^{\pi}\frac{{\rm d}q}{\pi}\,\kappa_q\,\sin^2q\,.
\end{eqnarray}
 The state (insulating or conducting) of the edge wire cannot affect bulk. Therefore, one has to first establish whether bulk is conducting ($\Delta_{\rm bulk}>1$) or insulating ($\Delta_{\rm bulk}<1$). Fixing the bulk state we have to test stability of the translation invariant edge wire against backscattering: if the scaling dimension ${\Delta}_{\rm edge}>1$, the edge wire is conducting and it is insulating otherwise. The phase of the system is governed by three parameters $0\leq K\leq\infty$ and $-1\leq\alpha_{\pm}\leq 1$. If both scaling dimensions $\Delta_{\rm bulk}$ and $\Delta_{\rm edge}$ are greater or smaller than one, we have complete conductor or insulator. If there existed a subspace where
\begin{equation}\label{ti}
\Delta_{\rm bulk} < 1 < \Delta_{\rm edge}\,,
\end{equation}
that would mean that we have situation similar to a quantum Hall system or a topological insulator: the bulk is insulating (low-temperature transport is blocked by backscattering from individual impurities in each wire) whereas edge wires are robust against backscattering generated by an impurity sitting at the edge. In the Fig.~(\ref{3D}) one can see that this situation arises when we have weak intra-channel attraction between fermions (or strong repulsion between bosons) and moderate inter-channel interactions.

In the Figure, one can see the result of numerical calculations. The plot of the scaling dimensions (vertical axis) for the bulk (blue surface) and the edge (yellow surface). The green plane is the guidance for the eye: points above it correspond to irrelevant perturbations with the scaling dimension above one. The region where the blue surface is below and the yellow surface is above green plane corresponds to the region Eq.~(\ref{ti}) of parameters ${\alpha}_{\pm}$ with conducting edge states and insulating bulk.

To ensure the edge state is 'protected' not only against a weak backscattering perturbation, one has to treat the opposite limit of a strong scatterer which is modelled by a cut with a weak tunneling across it (like in \cite{KF}). Using results on dualities in multi-channel Luttinger liquids we may claim that the scaling dimension $\Delta_{\rm edge}^{\rm tun}$ of the corresponding tunneling operator can be extracted from the generic result Eq.~(\ref{duality}):
\begin{equation}
\Delta_{\rm edge}^{\rm tun}=\Delta_{\rm edge}^{-1}
\end{equation}
This means that the edge wire can be in either conducting or insulating phase since scaling dimensions of perturbations above these phases are inversely proportional to each other. The absence of a mixed state corresponding to an unstable fixed point (which would be possible only if the product $\Delta_{\rm edge}^{\rm tun}\,\Delta_{\rm edge}$ was greater then one) leads to conclusion that the edge state, if conducting, is robust against both weak and strong scattering and, therefore, 'protected' by interaction at low-temperatures.

\section{Conclusions}
We have developed a universal approach to calculation of scaling dimensions of generic scattering operators in multi-channel Luttinger liquids in arbitrary configuration, i.e when some of them ideal (conducting) and others are interrupted (insulating). The problem was reduced to a single matrix equation. The approach allowed to analyse a finite array of interacting Luttinger liquids and define set of parameters where insulating bulk coexists with conducting edges.
The situation is similar to the one observed in quantum Hall systems and topological insulators. The model under consideration requires neither magnetic field nor spin-orbit interaction to observe quantised low-temperature edge transport.

%\begin{figure}\label{3D}
%\centerline{ {\includegraphics[width=.4\textwidth]{3D.pdf}}}
%\caption{\label{Fig2} On horizontal axes we have inter-channel strength parameters ${\alpha}_{\pm}$, the blue and green surfaces represent values of bulk and %edge scaling dimensions. The green plain is the eye guide for unity. The regions where the blue surface is under and the yellow surface above the green (unity) %one corresponds to insulating bulk and conducting edge state.}
%\end{figure}

\section*{Acknowledgement}The author thanks V. Kagalovsky for discussions and helpful comments on the manuscript. The hospitality of Center for Theoretical Physics of Complex Systems, Daejeon, South Korea, where the main part of this project was performed, is appreciated. This research was funded by the Leverhulme Trust Research Project Grant RPG-2016-044.

\appendix

\section{Translation invariant multi-channel LL}\label{AsLL}

First of all, one has to find the retarded Green function (we will suppress the superscript for 'retarded' component in what follows) corresponding to the ideal (translation invariant) multi-channel Luttinger liquid model (see Lagrangian Eq.~(\ref{L0})) using transformation Eq.~(\ref{pm}) from the next \ref{Vpm}:
\begin{eqnarray}\nonumber
{\hat G}_0(x;\omega)&=&2\,\int{\rm d}q\,\left(
                                            \begin{array}{cc}
                                              -{\hat V}_+q^2 & \omega q \\
                                              \omega q & -{\hat V}_-q^2 \\
                                            \end{array}
                                          \right)^{-1}\,e^{iqx}\\
&=&{\hat{\cal M}}\,{\hat g}_0(x;\omega)\,{\hat{\cal M}}^{\rm T}\,.
\end{eqnarray}
Here ${\hat g}_0$ is diagonal in channel indices Green function that describes set of $N$ uncorrelated one-dimensional Luttinger liquids with renormalised (by interactions) velocities:
\begin{eqnarray}\label{gf0}\nonumber
{\hat g}_0(x;\omega)&=&2\,\int{\rm d}q \,\left(
                                            \begin{array}{cc}
                                              -{\hat v}q^2 & \omega q \\
                                              \omega q & -{\hat v}q^2 \\
                                            \end{array}
                                          \right)^{-1}\,e^{iqx}\\
&=&-\frac{2\pi i}{\omega}\left(
                                            \begin{array}{cc}
                                              {\hat 1} & {\rm sgn}(x) \\
                                              {\rm sgn}(x) & {\hat 1} \\
                                            \end{array}
                                          \right)\,e^{i{\hat\omega}|x|}\,.
%\end{array}
\end{eqnarray}
The transformation is performed in density and current sectors without mixing them:
\begin{equation}
{\hat{\cal M}}=\left(
           \begin{array}{cc}
             {\hat M} & 0 \\
             0 & {\hat M}^{-\rm T} \\
           \end{array}
         \right)\,.
\end{equation}
We have also introduced the notations ${\hat\omega}=\frac{\omega}{{\hat v}}$ for the channel dependent frequency renormalisation. The ideal multi-channel LL is described by the following Green function
\begin{equation}\label{AGF0}
G_0(x;\omega)=-\frac{2\pi i}{\omega}\left(
                                            \begin{array}{cc}
                                              {\hat K} & {\rm sgn}(x) \\
                                              {\rm sgn}(x) & {\hat K}^{-1} \\
                                            \end{array}
                                          \right)\,e^{i{\hat\omega}|x|}\,,\,\,
{\hat K}={\hat M}\,{\hat M}^{\rm T}\,.
\end{equation}
where we introduced the Luttinger ${\hat K}$-matrix derived in the \ref{Vpm} and introduced in the main text Eq.~(\ref{K}).

If we needed equal coordinate correlation functions only (let us say at $x=0$), we could work out the corresponding effective Lagrangian. At zero-temperature we would have to invert the causal local Green function ($G^{\rm c}=-2\pi\,i/|\omega|$) to come to the result:
\begin{equation}
{\cal L}_{\rm ideal}=i\int\frac{{\rm d}\omega}{2\pi}\,|\omega|\left[{\bm \theta}^{\dagger}_{\omega}\,{\hat K}^{-1}\,{\bm \theta}_{\omega}+
{\bm \phi}^{\dagger}_{\omega}\,{\hat K}\,{\bm \phi}_{\omega}\right]
\end{equation}

\section{Lagrangian diagonalisation for translation invariant problem}\label{Vpm}
Given two real (all matrices in this Appendix will be real) symmetric positive definite matrices (the positiveness is required for stability)
\begin{equation}
{\hat V}_{\pm}={\hat V}_{\pm}^{\rm T}\,,\quad {\hat V}_{\pm}\succ 0\,,
\end{equation}
we have uniquely defined square roots because positive definite matrices have positive eigenvalues:
\begin{eqnarray}
{\hat V}&=&{\hat V}_{\pm}^{\rm T}\,,\quad {\hat V}\succ 0\,,\\
{\hat V}&=&{\hat O}\,{\hat\lambda}\,{\hat O}^{\rm T}\,,\quad {\hat O}\,{\hat O}^{\rm T}=1\,\quad \lambda_i \geq 0\,,\\
{\hat V}^{1/2}&=&{\hat O}\,{\hat\lambda}^{1/2}\,{\hat O}^{\rm T}\,.
\end{eqnarray}
The product (which is a non-symmetric matrix) of any two positive definite matrices has also positive eigenvalues:
\begin{equation}
{\hat V}_{-}\,{\hat V}_{+}={\hat V}_{-}^{1/2}\left[{\hat V}_{-}^{1/2}{\hat V}_{+}\,{\hat V}_{-}^{1/2}\right]{\hat V}_{-}^{-1/2}
\sim \, {\hat V}_{-}^{1/2}{\hat V}_{+}\,{\hat V}_{-}^{1/2} \succ 0\,.
\end{equation}
Since the product is similar to a positive definite matrix (the last one in the equation above), they have identical eigenvalues, i.e. positive ones. Lets us call them $v_i^2$ and the diagonal matrix of eigenvalues ${\hat v}^2=\textrm{diag}(v_1^2, v_2^2, ..., v_N^2)$. Then we can diagonalise this product by similarity transformation with some matrix ${\hat M}$:
\begin{equation}\label{product}
{\hat V}_{-}\,{\hat V}_{+}={\hat M}\,{\hat v}^2\,{\hat M}^{-1}\,.
\end{equation}
It turns out that one can solve this equation for ${\hat V}_+$ and ${\hat V}_-$. To do it, let us write equation transpose to Eq. (\ref{product})
\begin{equation}
{\hat V}_+\,{\hat V}_-={\hat M}^{-\rm T}\,{\hat v}^2\,{\hat M}^{\rm T}\,,
\end{equation}
then find ${\hat V}_+$ from the last equation and plug it back into Eq. (\ref{product}). The resultant equation will have the following form
\begin{equation}
{\hat\Lambda}_-\,{\hat v}^2={\hat v}^2\,{\hat\Lambda}_-\,,\quad
{\hat\Lambda}_-={\hat M}^{-\rm T}\,{\hat V}_-\,{\hat M}^{-1}\,.
\end{equation}
The only solution of this equation is $\Lambda_-$ which is a diagonal matrix. Repeating the same trick with ${\hat V_+}$ we find similar expression:
 \begin{equation}
{\hat\Lambda}_+\,{\hat v}^2={\hat v}^2\,{\hat\Lambda}_+\,,\quad
{\hat\Lambda}_+={\hat M}\,{\hat V}_+\,{\hat M}^{\rm T}\,.
\end{equation}
and similar conclusion: $\Lambda_-$ is a diagonal matrix. As the result, we can claim that
\begin{eqnarray}
{\hat V}_-&=&{\hat M}\,{\hat\Lambda}_-\,{\hat M}^{\rm T}\,,\\
{\hat V}_+&=&{\hat M}^{-\rm T}\,{\hat\Lambda}_+\,{\hat M}^{-1}\,.
\end{eqnarray}
To provide consistency with the Eq. (\ref{product}), we have to demand
\begin{eqnarray}
{\hat\Lambda}_-\,{\hat\Lambda}_+={\hat v}^2\,.
\end{eqnarray}
There is a freedom here, but we would like to have both $\Lambda$'s equal (${\hat\Lambda}_-={\hat\Lambda}_+={\hat v}$) because then they are interpreted as velocities in the new channels. Our final result:
\begin{eqnarray}\label{pm}
{\hat V}_-&=&{\hat M}\,{\hat v}\,{\hat M}^{\rm T}\,,\\
{\hat V}_+&=&{\hat M}^{-\rm T}\,{\hat v}\,{\hat M}^{-1}\,.
\end{eqnarray}
Please note that velocities are not eigenvalues of interaction matrices, their squares are eigenvalues of the products of interaction matrices. The important consequence of these two equations is that if they are written in the form
\begin{eqnarray}
{\hat v}={\hat M}^{-1}\,{\hat V}_-\,{\hat M}^{-\rm T}={\hat M}^{\rm T}\,{\hat V}_+\,{\hat M}\,\,,
\end{eqnarray}
then from the second equality one can get relation between two arbitrary symmetric positive-definite matrices
\begin{eqnarray}\label{KA}
{\hat K}\,{\hat V}_+\,{\hat K}={\hat V}_-\,,
\end{eqnarray}
that may be treated as the equation for the Luttinger matrix ${\hat K}$ which was defined as
\begin{equation}\label{k}
{\hat K}={\hat M}\,{\hat M}^{\rm T}\,,
\end{equation}
and, therefore, is real symmetric positive definite $N\times N$ matrix. This completes the proof of our main Eq.~(\ref{K}).

\section{Solutions to Eq.~(\ref{K}): Luttinger ${\hat K}$-matrix}\label{apps}
The Eq.~(\ref{K}) can be easily solved in the two limits: two-channel LL model and infinite array of 1D wires. These two applications are discussed in this Appendix.

\subsection{2-channel problem}
The equation can be solved using Cayley-Hamilton theorem for the matrix of velocities (or similar to it matrix $\sqrt{{\hat V}_-{\hat V}_+}$, see Eq.(\ref{product})):
\begin{equation}\label{CH}
{\hat v}^2-t\,{\hat v}+d=0\,,
\end{equation}
Here the coefficients of the characteristic polynomial are
\begin{equation}
t={\rm tr}{\hat v}={\rm tr}\sqrt{{\hat V}_-{\hat V}_+}\,,\quad d={\rm det}{\hat v}={\rm det}^{1/2}{\hat V}_-{\hat V}_+\,.
\end{equation}
Using Eqs.~(\ref{k}) and (\ref{CH}), we can write
\begin{equation}\label{CH1}
{\hat K}={\hat M}\,{\hat M}^{\rm T}={\hat M}\,\frac{1}{t}\,\left[{\hat v}+d\,{\hat v}^{-1}\right]{\hat M}^{\rm T}
=\frac{1}{t}\,\left[{\hat V}_-+d\,{\hat V}_+^{-1}\right]\,.
\end{equation}
To write explicit expression for the elements of the ${\hat K}$-matrix in terms of $V_{\pm}^{ij}$ one can use identity
\begin{equation}
\left({\rm tr}{\hat q}\right)^2={\rm tr}{\hat q}^2+2\,{\rm det}^{1/2}{\hat q}^2
\end{equation}
valid for any $2\times 2$ matrix and use it then for ${\hat q}=\sqrt{{\hat V}_-\,{\hat V}_+}$. The explicit expression follows immediately:
\begin{equation}
{\hat K}={\kappa}^{1/2}\,\left(
                           \begin{array}{cc}
                             \gamma\,\cosh\beta & \sinh\beta \\
                             \sinh\beta & \gamma^{-1}\,\cosh\beta \\
                           \end{array}
                         \right)\,.
\end{equation}
Here the parameters of the ${\hat K}$-matrix are related to the matrix elements of the interaction matrices:
\begin{eqnarray}
\kappa &=&{\det}^{1/2}\frac{{\hat V}_-}{{\hat V}_+}\,,\quad u_{ij}=V^{ii}_-+\kappa\,V^{jj}_+\,,\\
\tanh\beta &=&\frac{V_-^{12}-\kappa\,V_+^{12}}{\sqrt{u_{12}\,u_{21}}}\,,\quad \gamma^2=\frac{u_{12}}{u_{21}}\,.
\end{eqnarray}
This form of the ${\hat K}$-matrix has been used in \cite{KLY} and the corresponding phase diagram was constructed and analysed there.

\subsection{Infinite array}\label{sLL}
In an infinite ordered system, where all wires are parallel to each other and the perpendicular cross-section has wires arranged into $n$-dimensional lattice, with positions denoted as ${\bf r}$, we can introduce Fourier transform
\begin{eqnarray}
V_{{\bf r}-{\bf r}'}^{\pm}&=&\int\frac{{\rm d}^nq}{(2\pi)^n}\,V^{\pm}_{\bf q}\,e^{i{\bf q}({\bf r}-{\bf r}')}\,,\\
K_{{\bf r}-{\bf r}'}&=&\int\frac{{\rm d}^nq}{(2\pi)^n}\,K_{\bf q}\,e^{i{\bf q}({\bf r}-{\bf r}')}\,.
\end{eqnarray}
The Fourier transformation reduces Eq.~(\ref{K}) to the algebraic equation:
\begin{equation}
K_{\bf q}\,V^{+}_{\bf q}\,K_{\bf q}=V^{-}_{\bf q}\,,
\end{equation}
It is now trivial to return to the original representation and find matrix elements of the ${\hat K}$-matrix:
\begin{equation}
K_{{\bf r}-{\bf r}'}=\int\frac{{\rm d}^nq}{(2\pi)^n}\,K_{\bf q}\,e^{i{\bf q}({\bf r}-{\bf r}')}\,,\quad
K_{\bf q}=\sqrt{\frac{V^{-}_{\bf q}}{V^{+}_{\bf q}}}\,.
\end{equation}
This result can be found in early papers \cite{smectic}.

\section{Phase stability and Shortest Vector Problem}\label{SVP}
Any symmetric positive definite matrix can be written in the form of the Gram matrix, i.e. in the form when all matrix elements of the matrix are written as scalar products as $N$ linearly independent vectors ${\bf g}_i$:
\begin{equation}
K_{ij}={\bf g}^{\rm T}_i\,{\bf g}_j\,,\quad i,j=1\,,...\,, N\,.
\end{equation}
The inverse matrix is then defined by
\begin{equation}
\left[{\hat K}^{-1}\right]_{ij}={\bf a}^{\rm T}_i\,{\bf a}_j\,,\quad i,j=1\,,...\,, N\,,
\end{equation}
where sets $\left\{{\bf g}_i\right\}_{i=1}^N$ and $ \left\{{\bf a}_i\right\}_{i=1}^N\,,$
are dual to each other
\begin{equation}
{\bf a}_i^{\rm T}\,{\bf g}_i=\delta_{ij}.
\end{equation}
This construction is equivalent to the following parametrisation of the transformation matrix:
\begin{equation}
{\hat M}^{\rm T}=\left[{\bf g}_1\,,{\bf g}_2\,,...\,.{\bf g}_N\right]\,,\quad
{\hat M}^{-1}=\left[{\bf a}_1\,,{\bf a}_2\,,...\,.{\bf a}_N\right]\,.
\end{equation}
The scaling dimension of the most general perturbation for the most general (initially guessed/expected) configuration was derived in the main text:
\begin{equation}
\Delta[{\bf n}]={\bf n}^{\rm T}\,\left[{\hat{\cal P}}_{\rm c}\,{\hat K}^{-1}\, {\hat{\cal P}}_{\rm c}
+{\hat{\cal P}}_{\rm i}\,{\hat K}\,{\hat{\cal P}}_{\rm i}\right]^{-1}\,{\bf n}\,.
\end{equation}
Restricting ourselves to fully (i.e. all channels) conducting (${\hat{\cal P}}_i=0$) or insulating (${\hat{\cal P}}_c=0$) configuration, one has two bilinear forms built on integer-valued vectors
\begin{equation}
\Delta_{\rm cond}[{\bf n}]={\bf n}^{\rm T}\,{\hat K}\,{\bf n}\,,\quad \Delta_{\rm insul}[{\bf n}]={\bf n}^{\rm T}\,{\hat K}^{-1}\,{\bf n}\,,
\end{equation}
which should be minimised with respect to the integer-valued vector ${\bf n}$ (multiplicity of multi-particle scattering). The minimal value corresponds to the most dangerous term and its dimension should be then compared to the unity to see whether it is relevant or not and the configuration is stable (phase) or not. Using Gram representations (above)
\begin{eqnarray}
\min_{\bf n}\,\Delta_{\rm cond}[{\bf n}]&=&{\rm min}\,|{\bf G}|^2\,,\quad {\bf G}=\sum\limits_{i=1}^N\,n_i\,{\bf g_i}\,,\\
\min_{\bf n}\,\Delta_{\rm insul}[{\bf n}]&=&{\rm min}\,|{\bf R}|^2\,,\quad {\bf R}=\sum\limits_{i=1}^N\,n_i\,{\bf a_i}\,,
\end{eqnarray}
one can see that the problem is reduced to the shortest vector problem (SVP) \cite{SVP} on two mutually dual lattices, ${\bf G}\,{\bf R}={\rm integer}$. This representation relates conducting and insulating phases. The minimal distances between two points on two mutually dual lattices equal to scaling dimensions of perturbations above two phases of sliding Luttinger liquid.

The situation with inhomogeneous configuration can also be reduced to SVP:
\begin{equation}\label{nh}
\Delta[{\bf n}]=|{\bf G}_{\rm c}|^2+|{\bf R}_{\rm i}|^2\,,\quad
{\bf G}_{\rm c}=\sum_{j\subset{\rm {\bf C}}}\,n_j{\bf g}_j\,,\quad
{\bf R}_{\rm i}=\sum_{j\subset{\rm {\bf I}}}\,n_j{\bf a}_j\,.
\end{equation}
As it is clear from Eq.~(\ref{nh}), one has to define two mutually orthogonal subspaces spanned by vectors $\{{\bf a}\}_{j\subset{\rm {\bf C}}}$ (conducting channels) and $\{{\bf a}\}_{j\subset{\rm {\bf I}}}$ (insulting channels):
\begin{equation}
{\bf R}={\bf R}_{\rm c}+{\bf R}_{\rm c}\,,\quad {\bf R}_{\rm c}=\sum_{j\subset{\rm {\bf C}}}\,n_j{\bf a}_j\,,\quad
{\bf R}_{\rm i}=\sum_{j\subset{\rm {\bf I}}}\,n_j{\bf a}_j\,.
\end{equation}
Similarly, two subspaces must be formed for the dual lattice:
\begin{equation}
{\bf G}={\bf G}_{\rm c}+{\bf G}_{\rm c}\,,\quad {\bf G}_{\rm c}=\sum_{j\subset{\rm {\bf C}}}\,n_j{\bf g}_j\,,\quad
{\bf G}_{\rm i}=\sum_{j\subset{\rm {\bf I}}}\,n_j{\bf g}_j\,.
\end{equation}
The scaling dimension, Eq.~(\ref{Delta}), of any perturbation above arbitrary inhomogeneous configuration can be written as a sum, Eq.~(\ref{nh}). Therefore, to find scaling dimension one has to solve two SVP's on direct lattice corresponding to insulating channels and the dual sublattice corresponding to conducting ones. It is important to stress that these two subspaces are not independent because one should, first of all, construct the full lattice and its dual using the full Luttinger ${\hat K}$-matrix, and only after that split them up using insulating subspace if direct lattice and conducting subspace of the dual to it lattice.

\section{Duality}\label{duality}

Let us compare two arbitrary configurations different by only one element, i.e. some channel is assumed to be conducting in the first configuration and insulating in another one. We will call it first channel because we have not arranged channels spatially and the enumeration is irrelevant so far. According to the Eq.~(\ref{Delta}), one-particle backscattering in the first channel is governed by the scaling dimension
\begin{equation}
\Delta_{\rm c}=\left[\left({\hat{\hat{\cal P}}}_{\rm c}\,{\hat K}^{-1}\,{\hat{\hat{\cal P}}}_{\rm c}\right)^{-1}\right]_{11}\,,
\end{equation}
where the inversion is taken in the subspace of conducting channels. If the same wire was insulating, in the same environment of other channels, it would be disturbed by a tunneling with the scaling dimension given by
\begin{equation}
\Delta_{\rm i}=\left[\left({\hat{\hat{\cal P}}}_{\rm i}\,{\hat K}\,{\hat{\hat{\cal P}}}_{\rm i}\right)^{-1}\right]_{11}\,.
\end{equation}
Since in both cases the 'environment' is the same, we can introduce projection operators acting in $(N-1)\times (N-1)$ space of all channels except the first one:
\begin{equation}
{\hat{\hat{\cal P}}}_{\rm c}=\left(
                         \begin{array}{cc}
                           1 & 0 \\
                           0 & {\hat P}_{\rm c} \\
                         \end{array}
                       \right)\,,\quad
{\hat{\hat{\cal P}}}_{\rm c}=\left(
                         \begin{array}{cc}
                           1 & 0 \\
                           0 & {\hat P}_{\rm i} \\
                         \end{array}
                       \right)\,,\quad {\hat P}_{\rm c}+{\hat P}_{\rm i}=1\,.
\end{equation}
It is instructive to use ${\hat K}$-matrix representation in the block form separating conducting and insulating channels (see Eq. (\ref{block})):
\begin{equation}
{\hat K}=
\left(
  \begin{array}{cc}
    {\hat K}_{\rm cc} & {\hat K}_{\rm ci} \\
    {\hat K}_{\rm ic} & {\hat K}_{\rm ii} \\
  \end{array}
\right)\,.
\end{equation}
The channel one belongs to the ${\bf C}$-subspace in the one configuration and to the ${\bf I}$-subspace in the other. Since the projection of the inverse matrix can be calculated as
\begin{equation}
{\hat{\cal P}}_{\rm c}\,{\hat K}^{-1}\,{\hat{\cal P}}_{\rm c}=\left({\hat K}_{\rm cc}-{\hat K}_{\rm ci}{\hat K}_{\rm ii}{\hat K}_{\rm ic}\right)^{-1},
\end{equation}
we get the expression for the scaling dimension of the operator perturbing the first channel when it is in a conducting state:
\begin{equation}\label{c}
\Delta_{\rm c}=\left[{\hat K}_{\rm cc}-{\hat K}_{\rm ci}{\hat K}^{-1}_{\rm ii}{\hat K}_{\rm ic}\right]_{11}
= K_{11}-{\bf k}^{\rm T}_{\rm i}\,{\hat K}^{-1}_{\rm ii}\,{\bf k}_{\rm i}\,,
\end{equation}
where we have to define an auxiliary ${\bf k}_i$-vector
\begin{equation}
{\bf k}_{\rm i}={\hat P}_{\rm i}\,{\bf k}\,,\quad k^{\rm T}=\left(K_{12}\,, K_{13}\,, ...\,,K_{1N}\right)\,.
\end{equation}
Calculating alternative scaling dimension we need another projected matrix
\begin{equation}
{\hat{\hat{\cal P}}}_{\rm i}\,{\hat K}\,{\hat{\hat{\cal P}}}_{\rm i}=\left(
                                                                 \begin{array}{cc}
                                                                   K_{11} & {\bf k}^{\rm T}_i \\
                                                                   {\bf k}_i & {\hat K}_{\rm ii} \\
                                                                 \end{array}
                                                               \right)
\,.
\end{equation}
To find the inverse we may use cofactor matrix with $(1,1)$-minor given by ${\rm det}{\hat K}_{\rm ii}$:
\begin{equation}
\Delta_{\rm i}=\left[\left(
\begin{array}{cc}
K_{11} & {\bf k}^{\rm T}_i \\
{\bf k}_i & {\hat K}_{\rm ii} \\
\end{array}
\right)^{-1}\right]_{11}=\frac{{\rm det}\,{\hat K}_{\rm ii}}{{\rm det}({\hat{\hat{\cal P}}}_{\rm i}\,{\hat K}\,{\hat{\hat{\cal P}}}_{\rm i})
}\,.
\end{equation}
Then we can  the following reduction from $N\times N$ to $(N-1)\times (N-1)$ matrix representation:
\begin{equation}
{\rm det}\left({\hat{\hat{\cal P}}}_{\rm i}\,{\hat K}\,{\hat{\hat{\cal P}}}_{\rm i}\right)={\rm det}{\hat K}_{\rm ii}\,
{\rm det}\left[K_{11}-{\bf k}^{\rm T}_i\,{\hat K}^{-1}_{\rm ii}\,{\bf k}_i\right]\,.
\end{equation}
Now one can see that the scaling dimension of the perturbation above insulating state of the first channel is equal to
\begin{equation}\label{i}
\Delta_{\rm i}=\left[K_{11}-{\bf k}^{\rm T}_i\,{\hat K}^{-1}_{\rm ii}\,{\bf k}_i\right]^{-1}\,.
\end{equation}
The comparison of the results Eq.~(\ref{c}) and Eq.~(\ref{i}) proves the duality between scaling dimensions of perturbations in a single channel when it switches from conducting to insulating state:
\begin{equation}\label{duality}
\Delta_{\rm c}\,\Delta_{\rm i}=1\,.
\end{equation}
The duality equation is used in the text to prove that if all inner (bulk) wires are not switching their state, the edge wire can be in one of two states that are mutually exclusive, i.e. conducting and insulating phases of the edge exist in different parametric regions and never mix up, the unstable fixed point is impossible due to Eq. ~(\ref{duality}) and phases are robust against arbitrary strength of impurity scattering.

\end{document}